\documentclass[10pt,twocolumn,twoside]{IEEEtran}
\IEEEoverridecommandlockouts                           
\usepackage[noadjust]{cite}

\usepackage{amsmath,amssymb,amsfonts}
\usepackage{algorithmic}
\usepackage{graphicx}
\usepackage{textcomp}
\usepackage{balance}
\usepackage{xcolor}
\usepackage[utf8]{inputenc}
\usepackage{tikz}
\usepackage{booktabs}

\usetikzlibrary{shapes.geometric, arrows, positioning, fit, calc}

\def\BibTeX{{\rm B\kern-.05em{\sc i\kern-.025em b}\kern-.08em
    T\kern-.1667em\lower.7ex\hbox{E}\kern-.125emX}}

\title{\LARGE \bf 
Enhanced Optimal Power Flow Using a Trained Neural Network Surrogate for Distribution Grid Constraints}

\author{Savvas Panagi\IEEEauthorrefmark{1}\IEEEauthorrefmark{2}, 
Chrysovalantis Spanias\IEEEauthorrefmark{1}, 
and Petros Aristidou\IEEEauthorrefmark{2}\\
\IEEEauthorrefmark{1}Dept. of Electrical Computer Engineering and Informatics, Cyprus University of Technology, Limassol, Cyprus \\
\IEEEauthorrefmark{2}Distribution System Operator, Electricity Authority of Cyprus, Nicosia, Cyprus}

\begin{document}
\begingroup
\allowdisplaybreaks

\maketitle

\begin{abstract}
The growing penetration of distributed energy resources (DERs), electric vehicles (EVs), and heat pumps (HPs) in distribution networks underscores the need for secure, computationally efficient optimal power flow (OPF) solutions. Traditional OPF formulations often suffer from scalability limitations and may rely on relaxations/approximations whose \emph{exactness} is not guaranteed. This paper proposes a framework in which a trained neural network (NN) surrogate is embedded directly within the OPF as a constraint replacement. Specifically, the nonlinear power-flow-to-voltage mapping is replaced by an \emph{exact} mixed-integer linear encoding of the NN (i.e., the NN input--output map is represented without approximation), while all remaining OPF constraints are preserved. Using a realistic low-voltage network with integrated PV, EVs, and HPs, the proposed method achieves high voltage accuracy during post-solution AC power flow validation, with maximum deviations of less than 1.0~V in the examined test cases. The resulting NN-OPF problems are solved to global optimality within the MILP solver tolerance, and numerical results demonstrate substantially reduced computation time compared to nonlinear OPF models, with performance competitive with SOCP-based DistFlow formulations.
\end{abstract}

\begin{IEEEkeywords}
Neural Networks, Distribution Networks, Optimal Power Flow, Big-M Method, Surrogate Modeling
\end{IEEEkeywords}

%%%%%%%%%%%%%%%%%%%%%%%%%%%%%%%%%%%%%%%%%%%%%%%%%%%%%%%%%%%%%%%%%%%%%%%%%%%%%%%%

\section{Introduction}
\label{sec:introduction}

The transition towards Renewable Energy Sources (RES), together with the electrification of transportation and heating systems, has significantly increased the operational complexity of Distribution Networks (DNs). The Optimal Power Flow (OPF) problem plays a fundamental role in power system planning and operation, as it determines optimal operating points that minimize objectives such as generation cost or power losses while satisfying network and device constraints. However, the AC OPF formulation is inherently nonlinear and nonconvex, and is therefore classified as NP-hard \cite{bienstock2019strong}.

In modern DNs, OPF formulations must explicitly account for a wide range of controllable devices, including photovoltaic (PV) systems, Electric Vehicle (EV) charging stations, and Heat Pumps (HPs). As the penetration of such resources increases, the dimensionality and complexity of the OPF problem grow substantially. Consequently, there is a critical need for OPF formulations that can provide accurate and reliable solutions while remaining computationally efficient and scalable for operational and control applications.

% Traditional OPF formulations based on exact power flow equations, such as the full AC power flow provide accurate solutions but face computational challenges when applied to large-scale networks with many decision variables and constraints. 
% \textcolor{red}{defined the SOCP problems convex relaxations such as second-order cone programming (SOCP),} 

\subsection{Literature Review}
    Traditionally, the Bus Injection Model (BIM) has been widely adopted, expressing nodal power balances as nonlinear functions of voltage phasors. Despite its generality, the BIM formulation suffers from nonconvexity and poor scalability in large DNs \cite{molzahn2019survey}. A major shift followed with the introduction of the Branch Flow Model (BFM). The DistFlow formulation, a simplified case of the BFM for radial networks without shunt elements, originally proposed in \cite{baran1989optimal, baran2002optimal} for capacitor placement in DNs. Nevertheless, both BIM and BFM remain inherently nonconvex and, therefore, yield only locally optimal solutions.

In addition, several convexifications and approximations of the BFM have been proposed \cite{farivar2013branch, li2012exact}. Farivar and Low \cite{farivar2013branch} were among the first to show that an exact Second-Order Cone Programming (SOCP) relaxation of the BFM can yield globally optimal solutions under restrictive conditions, including radial topology, absence of shunt elements, and no lower bounds on generation. Building on this work, in \cite{li2012exact} derived broader sufficient conditions for exactness, demonstrating that the SOCP relaxation can remain exact even in the presence of voltage upper bounds and downstream generation, provided that certain conditions related to cumulative resistance-to-reactance ratios and downstream injections are satisfied. However, these conditions depend strongly on network structure and load–generation patterns and are rarely verified \emph{a priori} in practical distribution systems. Nevertheless, this line of research has motivated several extensions addressing shunt elements, line limits, and robustness \cite{christakou2017ac, zhou2020note}.

Neural network (NN)-based approaches have emerged as promising alternatives for accelerating power flow and OPF computations, with several representative implementations summarized in \cite{jiang2024advancements}. Most existing works focus on training neural networks to directly map system inputs, such as loads and generation, to complete OPF solutions \cite{nellikkath2022physics, baker2022emulating}. Other studies exploit learning-based control policies, including Markov decision processes and reinforcement learning, to bypass the classical OPF formulation altogether \cite{wang2022approximating, zhou2020data, sayed2022feasibility}.

More recently, hybrid approaches have been proposed to predict active constraint sets \cite{liu2021predicting, 8810819} or provide warm-start solutions \cite{baker2019learning}, thereby avoiding the need to fully replace the well-established, physically meaningful OPF formulation. While these methods can significantly accelerate optimization, they still suffer from fundamental limitations related to robustness, adaptability, and generalization. Recent work has explored the use of NN as surrogate models for specific components of the power flow problem \cite{simonovska2025electrical}. It has been argued that this targeted modeling represents one of the few practically viable approaches for operating low-voltage (LV) networks, where the reliance on detailed network models constitutes a major bottleneck for distribution system operators \cite{simonovska2022electrical}.

\subsection{Gap Analysis}

In practical distribution networks with high penetration of distributed generation (DG), frequent power flow reversals violate the assumptions required for the exactness of convex OPF relaxations, often leading to solutions that contradict the physical behavior of power flows \cite{andrianesis2021optimal}. Moreover, the sufficient conditions for SOCP exactness rarely hold in real-world networks due to highly heterogeneous line characteristics, including irregular and non-monotonic $r/x$ ratios, mixed underground and overhead segments, and limited controllability of nodal injections \cite{panagi2026thermalelectricalcooptimizationframeworkactive}. As a result, the practical feasibility of convex OPF relaxations must be carefully reassessed in flexible, DER-rich Active Distribution Networks (ADNs).

Direct NN-based OPF approaches face several fundamental limitations, including high computational complexity, limited adaptability, and weak generalization capabilities. Learning the full OPF solution space typically requires large training datasets, detailed physical network models, and complex NN architectures, information that is often unavailable to system operators due to data-access and privacy constraints. In addition, such approaches hinder flexibility, as incorporating new constraints or modifying the objective function generally requires retraining the network. Moreover, models trained on specific operating conditions may fail to generalize to unseen loading patterns, network topologies, or device configurations.

% Finally, to the best of our knowledge, existing NN surrogate models primarily focus on voltage estimation, while neglecting transformer and line loading constraints. Moreover, they typically predict the full network voltage profile, even though OPF formulations often require only the identification of worst-case voltages rather than all nodal values.

\subsection{Paper Contributions}

This work addresses the identified gaps through a novel modeling and optimization framework that replaces the nonlinear power flow equations with a trained neural network surrogate, while preserving the overall OPF structure and all remaining constraints. In contrast to direct NN-based OPF approaches, the proposed framework retains the mathematical optimization backbone, enabling flexibility, interpretability, and reliable optimality guarantees. The main contributions of this work are threefold:
\begin{itemize}
    \item We propose a data-efficient, model-free NN surrogate that learns only the voltage--power mapping, circumventing the feasibility and generalization issues of end-to-end NN-OPF methods while requiring substantially less training data, and preserving the OPF structure.
    \item We develop an exact MILP encoding of the NN via Big-M constraints that preserve the global optimality guarantees of the surrogate-constrained problem, unlike penalty- or projection-based feasibility-restoration schemes.
    \item We validate the framework on a realistic LV network with PV, EVs, and HPs, demonstrating sub-second solve times competitive with SOCP-DistFlow and near-optimal solutions with minimal voltage errors.
\end{itemize}

The remainder of this paper is organized as follows: Section~\ref{sec:opf} presents the general OPF formulation and introduces the constraint replacement strategy. Section~\ref{sec:nn_surrogate_milp} details the neural network surrogate model and its integration into the optimization framework using the Big-M method. Section~\ref{sec:case_study} presents the case study, Section~\ref{sec:results_discussion} discusses the results, and Section~\ref{sec:conclusion_future} concludes the paper.

\section{Optimal Power Flow Formulation and Constraint Replacement}
\label{sec:opf}

\subsection{General OPF Formulation}

The OPF problem for distribution networks with DGs, EVs, and HPs can be formulated as a multi-period optimization problem over a time horizon $\mathcal{T} = \{1, 2, \ldots, T\}$ \eqref{eq:gen_opt_prob}. 
\begin{subequations}
\label{eq:gen_opt_prob}
\begin{align}
\min_{\mathbf{x}} \quad & f(\mathbf{x}) \label{eq:optimization_problem_obj} \\
\text{s.t.} \quad & \mathbf{g}(\mathbf{x}) = \mathbf{0} \label{eq:optimization_problem_eq} \\
& \mathbf{h}(\mathbf{x}) \leq \mathbf{0} \label{eq:optimization_problem_ineq}
\end{align}
\end{subequations}
where $\mathbf{x}$ represents the decision variables including:
\begin{itemize}
    \item Power from controllable generation units: $P^{DG}_t$, $Q^{DG}_t$
    \item Charging power associated with the EV: $P^{EV}_t$, $Q^{EV}_t$
    \item Electrical and thermal power of HP: $P^{HP}_t$, $Q^{HP}_t$, $\mathbb{Q}^{HP}_t$
    \item Power exchange with the upstream grid: $P^{G}_t$, $Q^{G}_t$
    \item Network-related variables: $P^{L}_t$,$Q^{L}_t$, $V_t$, $I^L_t$
    \item EV energy and building temperature states: $E_t^{EV}$, $T^B_t$
\end{itemize}

The equality constraints $\mathbf{g}(\mathbf{x}) = \mathbf{0}$ include power flow consistency constraints, which enforce the electrical feasibility of the system by determining nodal voltages, branch power flows, and branch currents over the time horizon \eqref{eq:pf_mapping}. Moreover, it includes the electric vehicle battery charging dynamics \eqref{eq:soc_transition}, building temperature dynamics \eqref{eq:building_dynamics}, and HP operation \eqref{eq:hp_operation}. 
\begin{subequations}
\begin{align}
\big({V}_t, P^L_t, Q^L_t, I^L_t \big)
&= \mathcal{F}_{\mathrm{PF}}\left(P^{net}_t, Q^{net}_t, \mathbf{Y}\right),
\;\;\forall t \in \mathcal{T}
\label{eq:pf_mapping} \\
E^{EV}_t &= \mathcal{F}_{\mathrm{EV}}\left(P^{EV}_t\right),
\;\;\forall t \in \mathcal{T} 
\label{eq:soc_transition} \\
\mathbb{Q}^{HP}_t &= \mathcal{F}_{\mathrm{B}}\left(T^{B}_t\right),
\;\;\forall t \in \mathcal{T} 
\label{eq:building_dynamics} \\
P^{HP}_t &= \mathcal{F}_{\mathrm{HP}}\left(\mathbb{Q}^{HP}_t\right),
\;\;\forall t \in \mathcal{T} 
\label{eq:hp_operation} 
\end{align}
\end{subequations}
where $P^{net}_t$ and $Q^{net}_t$ represent net power injections at each bus (load minus generation), $\mathbf{Y}$ is the network admittance matrix.

The inequality constraints $\mathbf{h}(\mathbf{x}) \leq \mathbf{0}$ include:
\begin{itemize}
    \item Voltage limits: $\underline{V} \leq V_t \leq \overline{V}$
    \item DG power limits: $\underline{P}^{DG} \leq P^{DG}_t \leq \overline{P}^{DG}$, $\underline{Q}^{DG}\leq Q^{DG}_t \leq \overline{Q}^{DG}$
    \item EV charging constraints: $\underline{P}^{EV} \!\!\leq\!\!\ P^{EV}_t \!\!\leq \! \overline{P}^{EV}$, $\underline{E}^{EV}_t \!\!\leq\!\!\ E^{EV}_t \!\!\leq \! \overline{E}_t^{EV}$
    \item HP operational constraints and comfort limits: $\underline{P}^{HP} \leq P^{HP}_t \leq \overline{P}^{HP}$, $\underline{T}^{B}_t \leq T^{B}_t \leq \overline{T}^{B}_t$
\end{itemize}
The full formulation was presented in author's prior work \cite{panagi2026thermalelectricalcooptimizationframeworkactive}.
% Without reproducing the full formulation, which has been presented in authors' prior work
% the interested reader is referred to \cite{PSCC-paper1} 
% for a detailed description of the complete thermal–electrical co-optimization framework for active distribution networks with electric vehicles, heat pumps, and distributed generation.

\vspace{-1.5em}
\subsection{Constraint Replacement Strategy}

The key insight of the proposed approach is that the network power flow constraints in \eqref{eq:pf_mapping} constitute the most problematic component of the OPF problem, particularly when exact AC power flow formulations or their convex relaxations are employed. To address this bottleneck, these computationally intensive and often inexact constraints are replaced with a neural network surrogate. In doing so, the overall optimization structure is preserved: all decision variables, the objective function, and the remaining constraints remain unchanged. Moreover, the proposed approach retains the original high-level solution framework, including any decomposed or stakeholder-specific logic that may be subsequently integrated. The resulting modified OPF formulation is presented in \eqref{eq:mod_opt_prob}.
\begin{subequations}
\label{eq:mod_opt_prob}
\begin{align}
\min_{\mathbf{x}} \quad & f(\mathbf{x}) \label{eq:obj_nn} \\
\text{s.t.} \quad & \mathbf{h}(\mathbf{x}) \leq \mathbf{0} \label{eq:optimization_problem_ineq} \\
& V_t = \text{NN}(P^{net}_t, Q^{net}_t), \quad \forall t \in \mathcal{T} \label{eq:voltage_nn} \\
& \eqref{eq:soc_transition}, \eqref{eq:building_dynamics}, \eqref{eq:hp_operation}
\end{align}
\end{subequations}
where $\text{NN}$ represents the neural network surrogate model that will be formulated as MILP constraints in the next section.

\section{Neural Network Surrogate Integration using Big-M Method}
\label{sec:nn_surrogate_milp}

\subsection{Neural Network Architecture and Training}

The NN surrogate is designed to predict bus voltages as a function of net power injections. While the proposed formulation can, in principle, be extended to other quantities such as line currents or voltage phasors, voltage prediction is considered in this study, as voltage constraints constitute the primary operational bottleneck in LV distribution networks with high penetration of controllable loads. The NN architecture, depicted in Fig.~\ref{fig:neural_network_structure}, consists of the following layers:
\begin{itemize}
    \item \textbf{Input Layer}: Normalized net active and reactive power injections at each bus, $\mathbf{x}^{\mathrm{norm}} \in \mathbb{R}^{2N}$, where $N$ denotes the number of load buses.
    \item \textbf{Hidden Layer}: A fully connected layer with ReLU activation, parameterized by weights $\mathbf{W}_1 \in \mathbb{R}^{H \times 2N}$ and biases $\mathbf{b}_1 \in \mathbb{R}^{H}$, where $H$ is the number of hidden neurons.
    \item \textbf{Output Layer}: A linear layer that produces normalized voltage predictions, with weights $\mathbf{W}_2 \in \mathbb{R}^{N \times H}$ and biases $\mathbf{b}_2 \in \mathbb{R}^{N}$.
\end{itemize}
The forward pass of the neural network is given by:
\begin{subequations}
\label{eq:neural_network_equations}
\begin{align}
\mathbf{z} &= \mathbf{W}_1 \mathbf{x}^{norm} + \mathbf{b}_1 \label{eq:nn_hidden} \\
\mathbf{h} &= \text{ReLU}(\mathbf{z}) = \max(0, \mathbf{z}) \label{eq:nn_relu} \\
\mathbf{y}^{norm} &= \mathbf{W}_2 \mathbf{h} + \mathbf{b}_2 \label{eq:nn_output} \\
\mathbf{V} &= \mathbf{y}^{norm} \odot \boldsymbol{\sigma}^y + \boldsymbol{\mu}^y \label{eq:nn_denorm}
\end{align}
\end{subequations}
where $\boldsymbol{\mu}^{y}$ and $\boldsymbol{\sigma}^{y}$ denote the mean and standard deviation used for output denormalization, and $\odot$ represents element-wise multiplication. The NN is trained on data generated from power flow simulations across a wide range of operating scenarios, including varying load profiles, photovoltaic generation patterns, EV charging behavior, and HP operation.

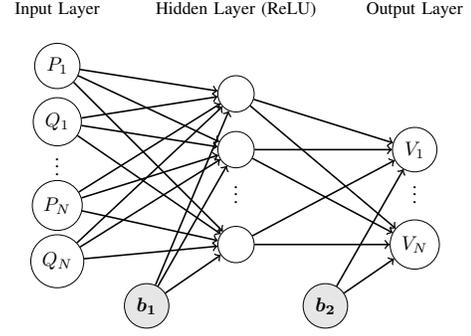
\begin{figure}
\centering
\resizebox{!}{0.5\columnwidth}{
\begin{tikzpicture}[
    neuron/.style={circle, draw, minimum size=6.5mm},
    bias/.style={circle, draw, minimum size=6.5mm, fill=gray!20},
    arrow/.style={->, thick},
    layerlabel/.style={font=\small}
]

% --------------------
% Input layer
% --------------------
\node[neuron] (P1) at (0,-1) {$P_1$};
\node[neuron] (Q1) at (0,-2) {$Q_1$};
\node at (0,-2.7) {$\vdots$};
\node[neuron] (PN) at (0,-3.5) {$P_N$};
\node[neuron] (QN) at (0,-4.5) {$Q_N$};

\node[layerlabel] at (0,0) {Input Layer};

% --------------------
% Hidden layer (wider spacing)
% --------------------
\node[neuron] (H1) at (3.2,-1.5) {};
\node[neuron] (H2) at (3.2,-2.5) {};
\node at (3.2,-3.2) {$\vdots$};
\node[neuron] (Hk) at (3.2,-4.2) {};

\node[layerlabel] at (3.2,0) {Hidden Layer (ReLU)};

% Bias (hidden)
\node[bias] (b1) at (1.6,-5.3) {$\boldsymbol{b_1}$};

% --------------------
% Output layer (wider spacing)
% --------------------
\node[neuron] (V1) at (6.4,-2.5) {$V_1$};
\node at (6.4,-3.2) {$\vdots$};
\node[neuron] (VN) at (6.4,-4.2) {$V_N$};

\node[layerlabel] at (6.4,0) {Output Layer};

% Bias (output)
\node[bias] (b2) at (4.8,-5.3) {$\boldsymbol{b_2}$};

% --------------------
% Connections Input -> Hidden
% --------------------
\foreach \i in {P1,Q1,PN,QN}
{
    \foreach \j in {H1,H2,Hk}
    {
        \draw[arrow] (\i) -- (\j);
    }
}

% Bias -> Hidden
\foreach \j in {H1,H2,Hk}
{
    \draw[arrow] (b1) -- (\j);
}

% --------------------
% Connections Hidden -> Output
% --------------------
\foreach \j in {H1,H2,Hk}
{
    \foreach \o in {V1,VN}
    {
        \draw[arrow] (\j) -- (\o);
    }
}

% Bias -> Output
\foreach \o in {V1,VN}
{
    \draw[arrow] (b2) -- (\o);
}

\end{tikzpicture}}
\caption{Neural network structure for voltage estimation.}
\label{fig:neural_network_structure}
\end{figure}

\subsection{MILP Formulation of Neural Network Constraints}

To integrate the NN into the optimization problem, we formulate \eqref{eq:neural_network_equations} as MILP constraints. The key challenge is the ReLU activation function, which is piecewise linear and non-differentiable. We utilize the Big-M method to precisely represent the ReLU function within the OPF framework.

% In the remainder, the term \emph{exact} refers to an exact mathematical encoding of the NN input--output relationship (and, in particular, the ReLU activation) as mixed-integer linear constraints, provided that the Big-M constants are valid bounds. This should not be confused with exactness of convex relaxations of AC power flow, nor does it imply that the NN surrogate reproduces AC power-flow voltages without error; surrogate accuracy is assessed separately via post-solution power flow validation.

The term \emph{exact} is used in a context-dependent manner and may refer either to an exact mathematical encoding of the NN input–output mapping as mixed-integer linear constraints, or to exactness with respect to power-flow formulations. The specific interpretation should be inferred from the context in which the term is used.

For each hidden neuron $j \in \{1, 2, \ldots, H\}$ and time step $t \in \mathcal{T}$, the pre-activation value $z_{j,t}$ is computed using the \eqref{eq:milp_hidden}.
\begin{align}
z_{j,t} = \sum_{i=1}^{2N} W_{1,ji} \cdot \frac{x_{i,t} - \mu^x_{i}}{\sigma^x_{i}} + b_{1,j} \label{eq:milp_hidden}
\end{align}
where input normalization uses mean $\boldsymbol{\mu}^x$ and standard deviation $\boldsymbol{\sigma}^x$ computed from the training data, and $x_{i,t}$ represents the $i$-th component of the net power injection vector at time $t$, computed as:
\begin{align}
\label{eq:input_component}
x_{i,t} = \begin{cases}
P^{net}_{k,t} & \text{if } i \leq N \text{ (active power at bus } k) \\
Q^{net}_{k,t} & \text{if } i > N \text{ (reactive power at bus } k)
\end{cases}
\end{align}
with $k = i$ for $i \leq N$ and $k = i - N$ for $i > N$. The net power at each bus accounts for loads, DGs, EVs, and HPs \eqref{eq:net_power}.
\begin{subequations}
\label{eq:net_power}
\begin{align}
P^{net}_{k,t} &= P^{load}_{k,t} - P^{DG}_{k,t} + P^{EV}_{k,t} + P^{HP}_{k,t}\\
Q^{net}_{k,t} &= Q^{load}_{k,t} - Q^{DG}_{k,t} - Q^{EV}_{k,t} + Q^{HP}_{k,t}
\end{align}
\end{subequations}

The ReLU activation $h_{j,t} = \max(0, z_{j,t})$ is reformulated using binary variables and the Big-M method. For each hidden neuron $j$ and time $t$, we introduce the continuous variable $h_{j,t} \geq 0$ representing the ReLU output, and a binary variable $\delta_{j,t} \in \{0,1\}$ indicating whether $z_{j,t} \geq 0$ (activated neuron).

The Big-M formulation uses a sufficiently large constant $M$ (typically $M = 1000$ based on the expected range of $z_{j,t}$) to enforce the following constraints. For any given valid bound $M$, this formulation exactly represents the ReLU function.
\begin{subequations}
\begin{align}
h_{j,t} &\geq z_{j,t} \label{eq:relu_lb1} \\
h_{j,t} &\geq 0 \label{eq:relu_lb2} \\
h_{j,t} &\leq z_{j,t} + M(1 - \delta_{j,t}) \label{eq:relu_ub1} \\
h_{j,t} &\leq M \delta_{j,t} \label{eq:relu_ub2} \\
z_{j,t} &\geq -M(1 - \delta_{j,t}) \label{eq:relu_z_lb} \\
z_{j,t} &\leq M \delta_{j,t} \label{eq:relu_z_ub}
\end{align}
\end{subequations}
These constraints ensure that:
\begin{itemize}
    \item If $z_{j,t} \geq 0$ (i.e., $\delta_{j,t} = 1$): Constraints (\ref{eq:relu_ub1}) and (\ref{eq:relu_z_ub}) become active, forcing $h_{j,t} = z_{j,t}$.
    \item If $z_{j,t} < 0$ (i.e., $\delta_{j,t} = 0$): Constraints (\ref{eq:relu_ub2}) and (\ref{eq:relu_z_lb}) become active, forcing $h_{j,t} = 0$.
\end{itemize}

The output layer computes normalized voltage predictions \eqref{eq:milp_output}, and then the actual voltage at each bus $k$ is calculated using \eqref{eq:vol_denorm}.
\begin{align}
y^{norm}_{k,t} &= \sum_{j=1}^{H} W_{2,kj} \cdot h_{j,t} + b_{2,k} \label{eq:milp_output} \\
V_{k,t} &= y^{norm}_{k,t} \cdot \sigma^y_{k} + \mu^y_{k} \label{eq:vol_denorm}
\end{align}

\subsection{Complete OPF Formulation}

The complete OPF problem with NN surrogate constraints is transformed into the following MILP:
\begin{subequations}
\label{eq:final_opf}
\begin{align}
\min_{\mathbf{x}, \mathbf{z}, \mathbf{h}, \boldsymbol{\delta}} \quad & f(\mathbf{x}) \label{eq:milp_obj} \\
\text{s.t.} \quad & \mathbf{h}(\mathbf{x}) \leq \mathbf{0} \label{eq:optimization_problem_ineq} \\
& \eqref{eq:soc_transition}, \eqref{eq:building_dynamics}, \eqref{eq:hp_operation} \\
& \text{Input and Hidden layer: } \eqref{eq:milp_hidden} - \eqref{eq:net_power} \\
& \text{ReLU activation: } (\ref{eq:relu_lb1})-(\ref{eq:relu_z_ub}) \\
& \text{Output layer: } (\ref{eq:milp_output}), (\ref{eq:vol_denorm}) \\
& \delta_j(t) \in \{0,1\} \quad \forall j, t
\end{align}
\end{subequations}

\subsection{Computational Considerations}

The Big-M constant $M$ should be chosen carefully. If it is too small, it may eliminate feasible solutions if $|z_j(t)| > M$ for some $j$ and $t$. If it is too large can lead to numerical issues and weak LP relaxations, slowing down the branch-and-bound algorithm. In practice, $M$ is selected based on the expected range of $z_j(t)$ values, which can be estimated from the training data or computed from the network weights and input bounds. For our implementation, we use $M = 1000$, which provides a good balance between feasibility and computational performance. The number of binary variables introduced is $H \times T$, where $H$ is the number of hidden neurons and $T$ is the number of time steps. While this increases problem complexity, modern MILP solvers (e.g., Gurobi) can efficiently handle problems with hundreds of binary variables.

\subsection{Simplified NN Surrogate for Minimum Voltage Constraint}

Since the operating decisions of interest are primarily driven by voltage limit violations, a simplified NN surrogate structure is also examined. Instead of predicting the full voltage profile across all buses, this formulation focuses exclusively on the minimum network voltage, which is the most critical indicator for feasibility under high load penetration. By restricting the NN output to the minimum voltage only, the resulting surrogate requires a significantly smaller NN architecture with fewer neurons, leading to a reduced number of binary variables in the MILP formulation and faster solution times, while still ensuring safe network operation. This simplified NN structure is hereafter referred to as Model 2, while the detailed NN formulation introduced earlier is denoted as Model 1.

\section{Case Study}
\label{sec:case_study}

\subsection{Distribution Network}

The proposed approach is validated on a realistic Cyprus LV network located in a suburban area. The network comprises a substation equipped with an 11/0.4 kV, 315 kVA delta-wye transformer. Three radial feeders extend from the substation with a total of 146 nodes and 85 residential customers.

\subsection{Load and PV Generation Profiles}

For the basic household demand, annual consumption data from 68 residential households provided by the Cyprus distribution system operator (DSO) are used to construct the load profiles. These profiles are randomly assigned to the network loads. Photovoltaic generation is modeled using statistical installation data also provided by the DSO, which are used to randomly allocate rooftop PV capacities to residential buildings according to the corresponding probability mass function. Owing to the geographic proximity of the households, a common PV generation profile is assumed and scaled according to the installed PV capacity of each unit. The PV penetration level considered in this study is 50\%.

\subsection{Electric Vehicle and Heat Pump}
Arrival and departure times, as well as travel distances, are derived from a survey conducted among car owners in Cyprus who rely on private vehicles for daily commuting \cite{sokr_pissa}. Furthermore, probability distribution functions based on the survey data are used to generate, for each EV, the corresponding arrival and departure times, as well as travel distances. For building temperature preferences, the same arrival and departure time distributions are adopted to generate the occupancy temperature preferences. All the datasets and network modeling are further explained and given in \cite{panagi2025impact, panagi_2025_17369365}.

\subsection{Training Process}
The neural network surrogate is trained on 30,000 diverse operating scenarios generated through power flow simulations, covering a wide range of operating conditions. These scenarios are constructed by applying different scaling factors to the maximum power ratings of EVs and HPs, as well as to historical load and PV generation profiles, thereby capturing various levels of realistic network operating states. The NN is trained using the Adam optimizer with a learning rate of 0.001 and a mean squared error loss function. 

\subsection{Optimization Problem and Scenarios}

The objective function considered in this case study minimizes the operating cost based on day-ahead electricity prices, $f(\mathbf{x}) = c_t \cdot \mathbf{x}$. While cost minimization is adopted here for illustration, it is not the primary focus of this work; the proposed framework is general and can readily accommodate alternative objectives, such as loss minimization or other network-centric performance metrics. During the implementation, we compare the following four approaches:
\begin{enumerate}
    \item \textbf{SOCP-DistFlow-OPF} \cite{farivar2013branch}: The second-order cone programming relaxation of the DistFlow model.
    \item \textbf{Nonlinear-OPF} \cite{peschon1972optimal}: The nonlinear AC optimal power flow formulation based on the bus injection model.
    \item \textbf{NN-OPF (Model 1)}: Proposed neural-network-based OPF (NN-OPF) formulation incorporating NN surrogate constraints that approximate the full network voltage profile, using 20 neurons in the hidden layer.
    \item \textbf{NN-OPF (Model 2)}: Proposed NN-OPF formulation incorporating NN surrogate constraints that approximate only the minimum network voltage, using a reduced NN architecture with 5 neurons in the hidden layer.
\end{enumerate}

\section{Results and Discussions}
\label{sec:results_discussion}

The SOCP and MILP optimization problem was solved using the GUROBI solver, while the nonlinear formulation was handled using IPOPT. All implementations were carried out on a laptop equipped with an M4 Pro Max processor, featuring a 12-core CPU and 36~GB of unified memory. The scheduling problem was formulated over a single-day horizon with a 15-minute time resolution. All optimization models were implemented using the Pyomo framework.

\subsection{Solution Exactness and Voltage Violations}
\label{subsec:solution_exactness}

Post-solution AC power flow analysis is performed to assess the accuracy of the proposed NN surrogate models in reproducing network voltage behavior. The evaluation focuses on a representative operating condition with 30\% EV and HP penetration, which drives the system close to its voltage limits and constitutes a stressed operating scenario. The resulting performance metrics are summarized in Table~\ref{tab:metrics_sol_exact_summary}, while the corresponding minimum voltage profiles are illustrated in Fig.~\ref{fig:min_voltage_comp}.

The results indicate that Model 1 consistently outperforms Model~2 in terms of minimum voltage accuracy across all evaluated metrics. In particular, Model 1 exhibits lower errors and improved robustness under stressed operating conditions. Notably, its root mean squared error (RMSE) remains comparable to the values observed during the NN testing phase, with voltage deviations remaining below 0.25~V, which is considered acceptable for practical applications. Even in the worst-case scenario, the maximum voltage deviation does not exceed approximately 0.4~V. Model~2 also demonstrates satisfactory performance, with the corresponding RMSE and maximum voltage deviation remaining below 0.4~V and 1.0~V, respectively.

In addition to minimum voltage evaluation, the voltage prediction accuracy of Model~1 across all buses is also reported in Table~\ref{tab:metrics_sol_exact_summary}. The overall RMSE across the network remains below 0.12~V, while the maximum observed deviation does not exceed 0.5~V. As shown in Fig.~\ref{fig:min_voltage_comp}, voltage limit violations mainly occur during high loading periods, particularly between 18:00 and 22:00. Under these conditions, four voltage violations are observed for Model~1 and three for Model~2. However, the minimum voltage values remain above 0.949~p.u., indicating that these deviations fall within the expected approximation error of the NN model and are not considered critical. The per-bus RMSE distribution is illustrated in Fig.~\ref{fig:rmse_per_bus}.

\begin{figure}
    \centering
    \includegraphics[width=1\linewidth]{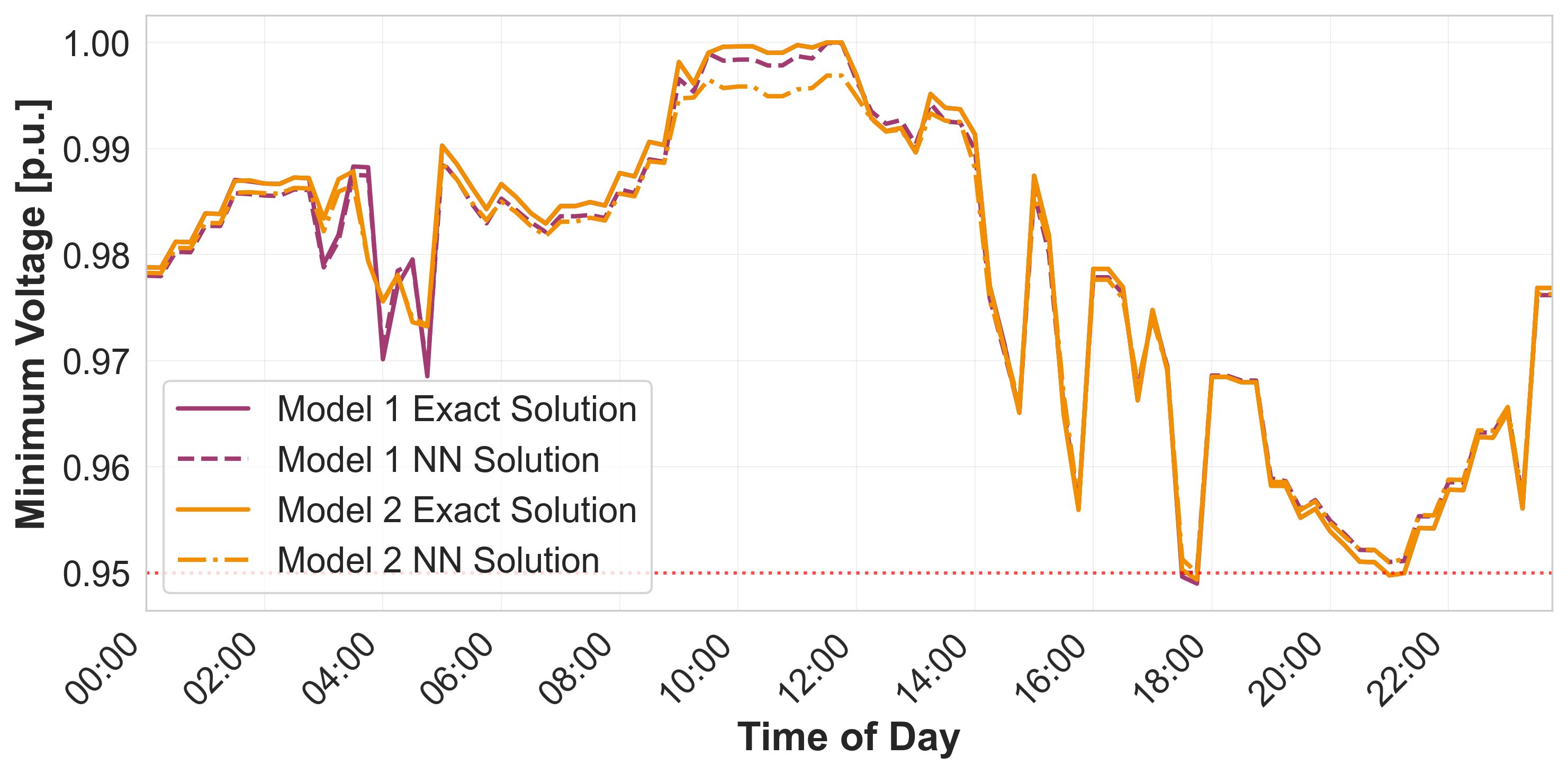}
    \caption{Comparison of minimum network voltage profiles for Model 1 and Model 2 over a 24-hour period.}
    \label{fig:min_voltage_comp}
\end{figure}

\begin{table}
    \centering
    \setlength{\tabcolsep}{4pt}
    \caption{Prediction accuracy metrics of the NN surrogates for voltage estimation, comparing the full voltage profile and minimum-voltage-based formulations.}
    \label{tab:metrics_sol_exact_summary}
    \begin{tabular}{lccc}
        \toprule
        \textbf{Metric} & \textbf{\begin{tabular}{c} Model 1 \\ (All Buses) \end{tabular}} & \textbf{\begin{tabular}{c}
        Model 1 \\ (Min Voltage)
        \end{tabular}} & \textbf{\begin{tabular}{c}
        Model 2 \\ (Min Voltage)
        \end{tabular}} \\
        \midrule
        RMSE (p.u.) & 0.000472 & 0.000992 & 0.001628 \\
        MAE (p.u.) & 0.000355 & 0.000899 & 0.001264 \\
        Max Error (p.u.) & 0.002120 & 0.001664 & 0.004178 \\
        MAPE (\%) & 0.0359 & 0.0917 & 0.1284 \\
        R\textsuperscript{2} & 0.998 & 0.996 & 0.988 \\
        \bottomrule
    \end{tabular}
\end{table}

\begin{figure}
    \centering
    \includegraphics[width=1\linewidth]{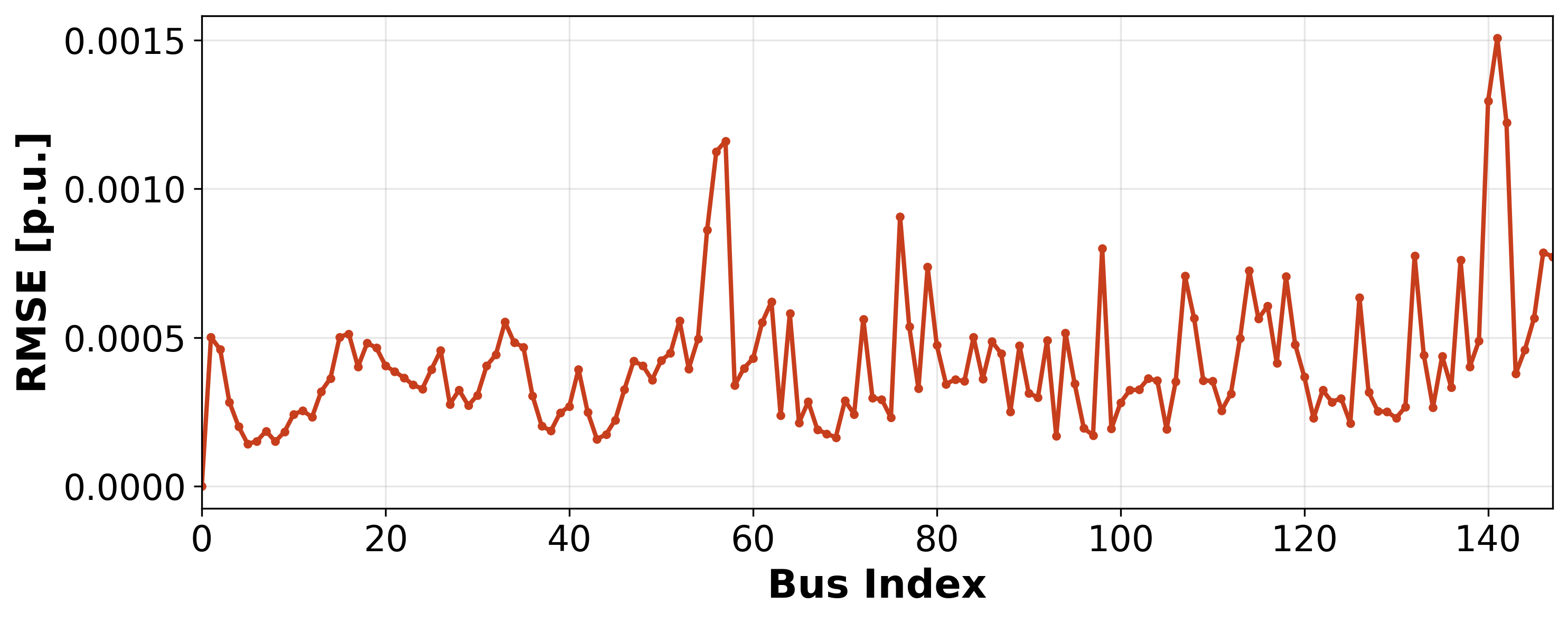}
    \caption{RMSE of the predicted bus voltages for NN-Model 1 across all network buses.}
    \label{fig:rmse_per_bus}
\end{figure}

\subsection{Computational Complexity and Scalability Robustness of the Proposed Formulations}

To assess the computational burden and practical scalability of the proposed OPF formulations, a robustness analysis is conducted across the three representative power flow formulations. The results in Table \ref{tab:opf_comparison} reveal a clear trend in the computational burden of the three OPF formulations as the penetration of EVs and HPs increases. 

The classic BIM formulation consistently exhibits the highest computational burden, with solution times increasing sharply from 39~s at 0\% penetration to 313~s at 45\% penetration. This steep growth indicates poor scalability under high active load penetration, limiting its practicality for large-scale or real-time applications. In contrast, the convex DistFlow formulation demonstrates significantly improved scalability, with computation times increasing only marginally from 1.2~s to 2.1~s over the same penetration range.

The proposed NN-based formulations provide an efficient alternative. Model~1 exhibits slightly higher solution times than the DistFlow model under low to moderate penetration levels, while a noticeable increase in computation time is observed at penetration levels exceeding 45\%. Nevertheless, solution times remain within the sub-second range, which is acceptable for most operational applications. In contrast, Model~2 consistently achieves the lowest computation times, ranging from 0.5~s to 1.2~s, outperforming even the convex DistFlow formulation.

\begin{table}
\setlength{\tabcolsep}{4pt}
\caption{Runtime [s] / objective value for OPF formulations under increasing EVs and HPs penetration.}
\label{tab:opf_comparison}
\centering
\begin{tabular}{c|c|c|c|c}
\hline
\textbf{\begin{tabular}{c}
EV+HP \\ Penetration
\end{tabular}}
& \textbf{BIM} 
& \textbf{DistFlow} 
& \textbf{\begin{tabular}{c}
NN \\ Model 1
\end{tabular}}
& \textbf{\begin{tabular}{c}
NN \\ Model 2
\end{tabular}} \\
\hline
0\%   & 38.6/0.0  & 1.2/0.0 & 2.4/0.0  & 0.5/0.0 \\ \hline
15\%   & 161.8/57.0  & 1.6/55.6 & 4.3/55.6  & 0.7/55.7 \\ \hline
30\%   & 158.0/63.0  & 1.7/61.9 & 4.2/61.9  & 0.7/61.9 \\ \hline
45\%   & 312.7/89.1  & 2.1/85.1 & 18.7/85.1  & 1.2/85.3 \\ \hline
\end{tabular}
\end{table}

\subsection{Solution Quality}

Post-solution analysis indicates that the SOCP relaxation is exact for the examined scenarios. Consequently, the SOCP solution is treated as the reference global optimum (for the relaxed formulation) against which all objective values are compared. The objective values obtained with each formulation are reported in Table~\ref{tab:opf_comparison}. Both NN-based formulations yield near-optimal objective values relative to this reference; note that NN-OPF (Models 1--2) are MILPs and are solved to global optimality within the solver's MIP gap tolerance, but they optimize a surrogate-constrained problem rather than the original AC-OPF. The small deviations from the reference solution are primarily attributed to the solver's accepted optimality gap (1\%) and the minor voltage approximation errors discussed in Section~\ref{subsec:solution_exactness}. In contrast, the BIM formulation converges to a local optimum, yielding higher objective values than the reference global optimum.

\subsection{Sensitivity Analysis}

A sensitivity analysis is conducted to examine the impact of increasing the number of neurons on both computational time and solution accuracy for NN-OPF (Model~1). The results, illustrated in Fig.~\ref{fig:sens_model1}, indicate an approximately linear increase in solving time as the NN complexity grows with the number of neurons. At the same time, the prediction accuracy improves with the increase in neuron size in the hidden layer; however, the most significant improvement is observed up to approximately 30 neurons. Beyond this point, further increases in the number of neurons yield only marginal accuracy gains, while the computational cost continues to increase.

\begin{figure}
    \centering
    \includegraphics[width=1\linewidth]{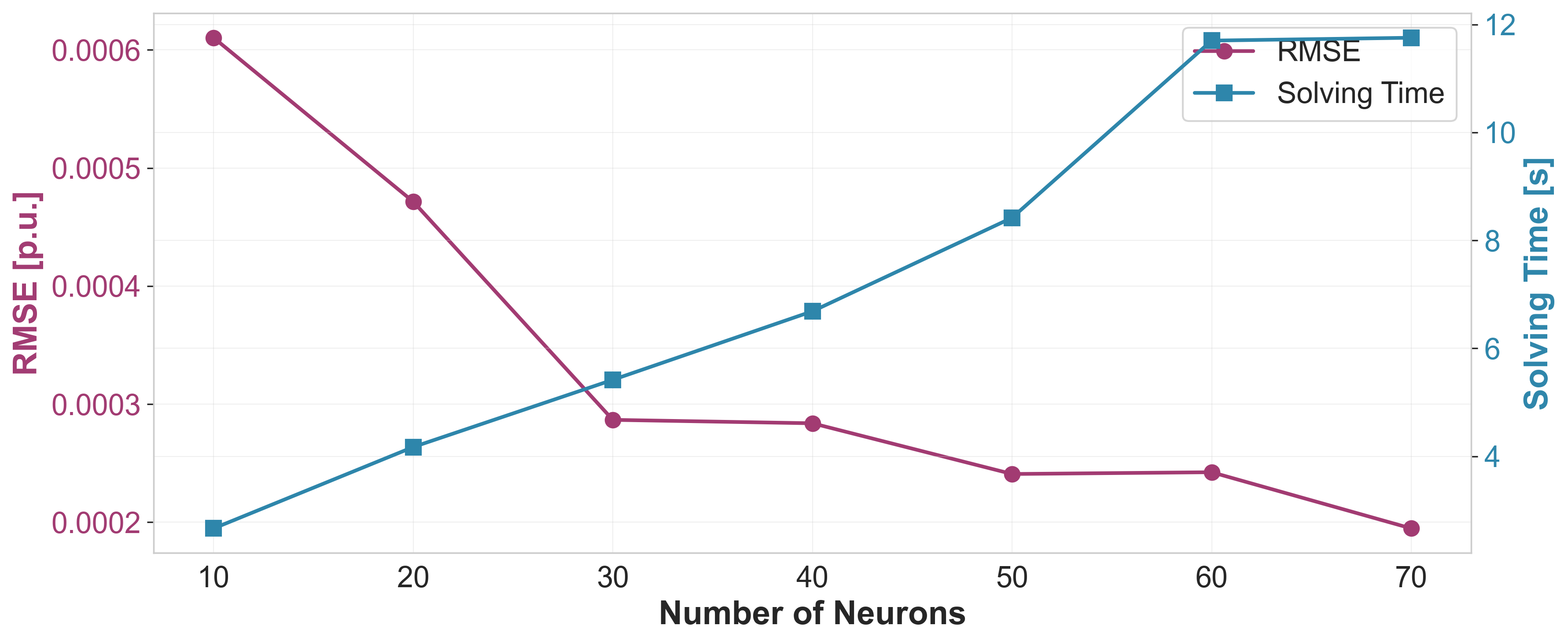}
    \caption{Trade-off between voltage prediction accuracy and computational time as a function of the number of neurons in the NN surrogate.}
    \label{fig:sens_model1}
\end{figure}

\subsection{Discussions}

% It's worth noting that the proposed method was initially tested on the CIGRE LV network, yielding results nearly identical to those of the SOCP, and with significantly faster solving times. However, due to the simplicity of this test network, which consists of only around 40 buses and a few controllable loads, the authors decided to move on to a more realistic and complex LV network in Cyprus, as provided by the electricity authority of Cyprus.

Based on these observations, Model~2 effectively combines low computational burden with strong optimization guarantees: despite the introduction of binary variables, the resulting surrogate-constrained problem can be solved to global optimality within MILP solver tolerances, while using a compact NN architecture. This makes it an attractive alternative to conventional convex relaxations, particularly in applications that require fast decision-making. Nevertheless, when detailed voltage information across all buses is necessary, Model~1 remains a viable and accurate option, albeit at a higher computational cost.

Most importantly, both NN-based formulations offer key advantages over SOCP-DistFlow, BIM, and other heuristic approaches. When trained on high-quality reference solutions, they consistently deliver high-voltage accuracy during post-solution power flow validation while retaining the transparency and flexibility of an optimization-based framework. As a result, the proposed NN surrogates can substantially reduce computation time while providing solutions that are near-optimal relative to the chosen reference benchmark in this case study.

\section{Conclusions and Future Works}
\label{sec:conclusion_future}

This paper presents a novel framework for integrating neural network surrogates into optimal power flow optimization for distribution networks. Unlike existing approaches that attempt to learn the complete OPF solution mapping, the proposed method selectively replaces only the most challenging component of the problem, namely, the nonlinear power-flow-to-voltage mapping, with a trained neural network, while preserving the underlying optimization structure and all remaining constraints. This targeted replacement retains the flexibility of the original OPF formulation, allowing for arbitrary objective functions and the straightforward inclusion of additional constraints. Moreover, the neural network is required to learn only the voltage--power relationship, which is substantially simpler and more robust. In the examined test case, the proposed framework achieves high voltage accuracy under post-solution AC power flow validation (maximum deviation below 1.0~V). The resulting NN-OPF formulations are solved to global optimality within MILP solver tolerances; this global optimality pertains to the surrogate-constrained optimization problem and should not be interpreted as global optimality for the original AC-OPF.

Despite these advantages, the proposed method currently focuses exclusively on voltage constraints and does not explicitly account for transformer or line loading limits. Additionally, relying on binary variables, while manageable for moderate-sized networks, may pose computational challenges for large-scale systems and extended time horizons. Consequently, future work will focus on incorporating branch and transformer loading estimation to enable a more comprehensive OPF formulation. Furthermore, alternative NN modeling approaches that avoid or significantly reduce the use of binary variables will be investigated, with the objective of enabling convex continuous formulations and further improving the global optimality guarantee.

\section*{Acknowledgment}
% AI assistance was used for code development (Cursor AI) and for syntactic-level editing of this manuscript (GPT-5.1).
\thanks{This project has received funding from the European Union’s Horizon Europe Framework Programme under Grant Agreement No. 101120278 (DENSE).}

% \newpage 
\bibliographystyle{ieeetr}
\bibliography{references}

\balance

\endgroup
\end{document}